\documentclass[twocolumn,showpacs,preprintnumbers,amsmath,amssymb]{revtex4}

\usepackage{graphicx}
\usepackage{dcolumn}
\usepackage{bm}
\usepackage{dsfont}
\usepackage{amsthm}
\usepackage{graphicx}

\theoremstyle{definition}
\newtheorem{Def}{Definition}[section]

\theoremstyle{plain}

\newtheorem{Thm}[Def]{Theorem}

\theoremstyle{remark}

\newtheorem*{Thm*}{}

\newcommand{\dd}{\mathrm{d}}
\newcommand{\ee}{\mathrm{e}}
\newcommand{\ii}{\mathrm{i}}
\newcommand{\rh}{r_\mathrm{h}}
\newcommand{\Hp}{{\mathcal H^+}}
\newcommand{\Hm}{{\mathcal H^-}}
\newcommand{\Hpm}{{\mathcal H^\pm}}
\newcommand{\E}{\mathcal E}
\newcommand{\f}{{\ee^{2U}}}

\newcommand{\C}{\mathds C}

\newcommand{\rW}{\varrho}
\newcommand{\zW}{\zeta}
\newcommand{\pmN}{^\pm_\mathrm{N}}
\newcommand{\pN}{^+_\mathrm{N}}

\newcommand{\pmS}{^\pm_\mathrm{S}}
\newcommand{\pS}{^+_\mathrm{S}}

\begin{document}


\title{Inner Cauchy horizon of axisymmetric and
 stationary black holes with surrounding matter in Einstein-Maxwell theory}

\author{Marcus Ansorg}
\email{mans@aei.mpg.de}
\author{J\"org Hennig}
\email{pjh@aei.mpg.de}
\affiliation{Max Planck Institute for Gravitational Physics,\\
Am M\"uhlenberg 1, D-14476 Golm, Germany}

\date{\today}

\begin{abstract}
 We study the interior electrovacuum region of axisymmetric and
 stationary black holes with surrounding matter and find 
 that there exists always a regular inner Cauchy horizon inside the
 black hole, provided the angular momentum $J$ and charge $Q$
 of the black hole do not vanish simultaneously. In particular, we derive an 
 explicit relation for the metric on the Cauchy
 horizon in terms of that on the event horizon.
 Moreover, our analysis reveals the remarkable universal relation
 $(8\pi J)^2+(4\pi Q^2)^2=A^+ A^-$, where 
 $A^+$ and $A^-$ denote the areas of event and Cauchy horizon,
 respectively.
\end{abstract}

\pacs{04.70.Bw, 04.40.-b, 04.40.Nr, 04.20.Cv}

\maketitle
\section{Introduction \label{sec:Intro}}

In the interior of a single rotating, electrically charged,
axisymmetric and stationary
black hole in electrovacuum (described by the Kerr-Newman family of
solutions) there exists a Cauchy horizon $\mathcal H^-$, which is
the future boundary of the domain of dependence of the event horizon
$\mathcal H^+$. The presence of this Cauchy horizon is related to the
fact that, in Boyer-Lindquist coordinates, the axisymmetric and stationary 
Einstein-Maxwell vacuum equations 
are hyperbolic within an interior vicinity of 
$\mathcal H^+$.
The two horizons
$\mathcal H^+$ and $\mathcal H^-$ represent the past and future boundary
of this hyperbolic region.
Interestingly, for the Kerr-Newman family 
the areas $A^\pm$ of the horizons $\mathcal H^\pm$
are related by
\begin{equation}\label{ApAm}
 (8\pi J)^2+(4\pi Q^2)^2 = A^+ A^-,
\end{equation}
where $J$ and $Q$ are angular momentum and charge of the black hole,
respectively. 

In pure Einsteinian gravity (i.e., without Maxwell field), 
these observations have been generalized in \cite{Ansorg2}. It
was shown that for axisymmetric and stationary black holes {\em with
surrounding matter} 
there always exists a regular inner Cauchy horizon, provided that $J\neq 0$
holds.
It was also shown that such black holes satisfy relation~\eqref{ApAm}
(with $Q=0$) in general.

In this Letter, we investigate axisymmetric and stationary black holes with
surrounding matter and include electromagnetic fields.
For our analysis, we will make use of a linear matrix problem,
whose integrability conditions are
equivalent to the Einstein-Maxwell equations in vacuum
(see \cite{Neugebauer}). The existence of such a linear problem (LP)
permits the application of so-called soliton methods 
(e.g.~the ``inverse scattering method'') through which
particular solutions of the field equations in question can be found
(see e.g.~\cite{Neugebauer2}, \cite{Neugebauer3}).

Here we integrate the LP along the boundaries of the
hyperbolic region inside the black hole and obtain in this manner useful
relations between the field quantities at these boundaries.
In particular, we are able to calculate the metric and electromagnetic
potentials on $\mathcal H^-$ in terms of those on $\mathcal
H^+$. Moreover, we find that from these relations Eq.~\eqref{ApAm} 
can be deduced; i.e., Eq. \eqref{ApAm} turns out to be valid for
arbitrary axisymmetric and 
stationary black holes with surrounding matter in Einstein-Maxwell theory.

A detailed description of the calculations sketched below
will be given in a forthcoming paper.

\section{Coordinate systems and Einstein-Maxwell equations
\label{sec:Coords}}

\begin{figure*}
 \centering
 \includegraphics[scale=0.8]{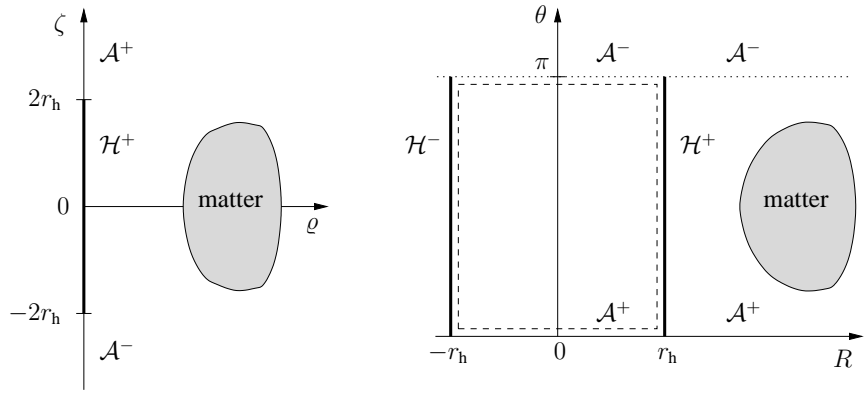}
 \caption{Sketch of a part of a black hole space-time
  in Weyl coordinates (\emph{left panel})
  and Boyer-Lindquist type coordinates (\emph{right panel}). $\mathcal
  A^+$ and $\mathcal A^-$ denote upper and lower parts of the
  symmetry axis and $\mathcal H^+$ and $\mathcal H^-$ denote event
  and Cauchy horizons. In Sec.~\ref{solution} we will integrate the linear
  problem for the Einstein-Maxwell equations along the closed dashed line.}
 \label{fig1}
\end{figure*}

In the following, we study an electrovacuum vicinity of the black hole's event
horizon. (Note that for a stationary spacetime, the
immediate vicinity of a black hole event horizon \emph{must} be
electrovacuum, see
\cite{Carter} and \cite{Bardeen}.) In this vicinity, we introduce Weyl
coordinates $(\varrho, \zeta, \varphi, t)$ in which the line element reads as follows
\begin{equation}\label{LE1}
 \dd s^2 = \ee^{-2U}\left[\ee^{2k}(\dd\rW^2+\dd\zW^2)
            +\rW^2\dd\varphi^2\right]
           -\ee^{2U}(\dd t+a\dd\varphi)^2,
\end{equation}
where the metric potentials $U$, $k$, and $a$ are functions of $\varrho$
and $\zeta$ alone. As sketched in Fig.~\ref{fig1} (left panel),
the event horizon
$\Hp$ is located on the interval $-2\rh\le\zeta\le 2\rh$,
$\rh=\textrm{constant}$, of the $\zeta$-axis. The remaining part
$|\zeta|>2\rh$ of the $\zeta$-axis corresponds to the rotation
axis. 

In order to investigate the \emph{interior} of the black hole, which
is characterized by negative values of $\varrho^2$, we also introduce
Boyer-Lindquist-type coordinates $(R, \theta, \varphi, t)$ via
\begin{equation}\label{BoyerLcoord}
   \varrho^2 = 4(R^2-\rh^2)\sin^2\!\theta,\qquad
   \zeta  = 2R\cos\theta.
\end{equation}
In these coordinates, the event horizon $\Hp$ and the inner Cauchy
horizon $\Hm$ are located at $R=\rh$ and $R=-\rh$ respectively, see
Fig.~\ref{fig1} (right panel).

In the electrovacuum region,the electromagnetic field alone constitutes
the energy momentum tensor 
\begin{equation}
 T_{ij} = \frac{1}{4\pi}\left(F_{ki}F^k_{\ j}
          -\frac{1}{4}g_{ij}F_{kl}F^{kl}\right),
\end{equation}
where $F_{ij}$ is the electromagnetic field tensor.
We use the Lorenz gauge, in which $F_{ij}$ can be written in terms of a
vector potential $(A_i) = (0, 0, A_{\varphi}, A_t)$,
\begin{equation}
 F_{ij}=A_{i,j}-A_{j,i}.
\end{equation}
Note that, like the metric quantities, $A_\varphi$ and $A_t$ also depend on $R$
and $\theta$ only.

We introduce the complex electromagnetic potential $\Phi$ and the complex
Ernst potential $\E$ \cite{Ernst, ExactSolutions} by
\begin{equation}\label{cpot}
 \Phi = A_t + \ii\beta,\quad
 \E   =\ee^{2U}-|\Phi|^2+\ii b,
\end{equation}
where the imaginary parts $b$ and $\beta$ are related to metric and
vector potentials via
\begin{subequations}\label{ab1}
\begin{eqnarray}
 \label{a1}
 a_{,\varrho}  & = & \varrho\ee^{-4U}\left[b_{,\zeta}
                  -\ii(\bar\Phi\Phi_{,\zeta}-\Phi\bar\Phi_{,\zeta})\right],\\
 \label{a2}
 a_{,\zeta} & = & -\varrho\ee^{-4U}\left[b_{,\varrho}
                  -\ii(\bar\Phi\Phi_{,\varrho}-\Phi\bar\Phi_{,\varrho})\right],\\
 \label{b1}
 \beta_{,\varrho} & = &
 \frac{\ee^{2U}}{\varrho}(aA_{t,\zeta}-A_{\varphi,\zeta}),\\
 \label{b2}
  \beta_{,\zeta} & = &
 -\frac{\ee^{2U}}{\varrho}(aA_{t,\varrho}-A_{\varphi,\varrho}).
\end{eqnarray}
\end{subequations}

In this formulation (the bar denotes complex conjugation), the Einstein-Maxwell equations in electrovacuum are
equivalent to the two complex Ernst equations \cite{Ernst}
\begin{subequations}\label{EM}
\begin{eqnarray}
  \f\bigtriangleup\E & = & \nabla\E\cdot(\nabla\E+2\bar\Phi\nabla\Phi),\\
  \f\bigtriangleup\Phi & = & \nabla\Phi\cdot(\nabla\E+2\bar\Phi\nabla\Phi).
\end{eqnarray}
\end{subequations}
Here, $\bigtriangleup$ and $\nabla$ denote Laplace and nabla
operators in flat cylindrical coordinates $(\varrho,\zeta,\varphi)$.\\[.5ex]

\section{The linear problem}

As mentioned in the introduction, the existence of the LP, whose integrability conditions are
equivalent to \eqref{EM}, is crucial for our analysis. For its formulation we
introduce the complex coordinates
$z=\varrho + \ii\zeta$, $\bar z = \varrho - \ii\zeta$
and the function
\begin{equation}\label{la}
 \lambda(K,z,\bar z)=\sqrt{\frac{K-\ii\bar z}{K+\ii z}},
\end{equation}
which depends on a \emph{spectral parameter} $K\in\C$.
For fixed values $z$, $\bar z$, Eq. \eqref{la} describes a spectral
mapping $\C\to\C$,
$K\mapsto\lambda$ from a two-sheeted Riemann surface ($K$-plane)
onto the complex $\lambda$-plane. The two $K$-sheets are connected
at the two branch points $K_1=\ii\bar z$ ($\lambda=0$)
and $K_2=-\ii z$ ($\lambda=\infty$). 

The LP is a system of first order equations for a
$3\times 3$ matrix pseudopotential
${\bf\Omega}={\bf\Omega}(K,z,\bar z)$, which reads \cite{Neugebauer}
\begin{widetext}
\begin{subequations}\label{LP}
\begin{eqnarray}
 {\bf\Omega}_{,z} & = & \left[\left(\begin{array}{ccc}
                  B_1  & 0   & E_1\\
                  0    & A_1 & 0  \\
                  -F_1 & 0   & \frac{1}{2}(A_1+B_1)
                  \end{array}\right)
               + \lambda\left(\begin{array}{ccc}
                  0    & B_1 & 0\\
                  A_1  & 0   & -E_1\\
                  0    & -F_1& 0
                  \end{array}\right)\right]{\bf\Omega},\\[0.3ex]
 {\bf\Omega}_{,\bar z} & = & \left[\left(\begin{array}{ccc}
                  B_2  & 0   & E_2\\
                  0    & A_2 & 0  \\
                  -F_2 & 0   & \frac{1}{2}(A_2+B_2)
                  \end{array}\right)
               + \frac{1}{\lambda}\left(\begin{array}{ccc}
                  0    & B_2 & 0\\
                  A_2  & 0   & -E_2\\
                  0    & -F_2& 0
                  \end{array}\right)\right]{\bf\Omega},            
\end{eqnarray}
\end{subequations}
\end{widetext}
where
\begin{subequations}\label{ABEF}
\begin{eqnarray}
& A_1 =  \frac{1}{2}\ee^{-2U}(\E_{,z}+2\bar\Phi\Phi_{,z}),\quad
 E_1 = \ii\ee^{-U}\Phi_{,z},\\[1.5ex]
& B_1 =  \frac{1}{2}\ee^{-2U}(\bar\E_{,z}+2\Phi\bar\Phi_{,z}),\quad
 F_1 = \ii\ee^{-U}\bar\Phi_{,z}.
\end{eqnarray}
\end{subequations}
Analogous expressions for
$A_2$, $B_2$, $E_2$, and $F_2$ can be obtained from
\eqref{ABEF} by replacing $z$ with $\bar z$.

If ${\bf\Omega}$ is a solution of the LP \eqref{LP}, then ${\bf\Omega
C}(K)$ is also a solution for every $3\times 3$ matrix function ${\bf C}(K)$.
We can 
always find a ${\bf C}(K)$ to bring ${\bf\Omega}$ into the form
\begin{equation}\label{gauge}
 {\bf\Omega}^>(K, z,\bar z)=\left(\begin{array}{crr}
              \psi_1^>(K,z,\bar z) &  \psi_1^<(K, z, \bar z) & 0\\
              \psi_2^>(K,z, \bar z) & -\psi_2^<(K, z, \bar z) & 0\\
              \psi_3^>(K,z, \bar z) &  \psi_3^<(K,z,\bar z) & 0
              \end{array}\right),
\end{equation}
which depends on three functions $\psi_1$, $\psi_2$, $\psi_3$. Here,
the superscripts ``$>$'' or ``$<$'' indicate whether the functions
are evaluated in the upper ($\lambda=1$ for $K=\infty$) or
lower ($\lambda=-1$ for $K=\infty$) sheet of the two-sheeted Riemann $K$-surface.
By interchanging ``$>$'' and ``$<$'' in \eqref{gauge}, we obtain a
similar equation for ${\bf\Omega}^<$. Obviously, ${\bf\Omega}$ in the form \eqref{gauge} is not
invertible. Nevertheless, we will see that it still contains sufficient
information about $\E$ and $\Phi$.

For our analysis of the LP it is also useful to study the
situation in a frame of reference that rotates with a constant
angular velocity $\omega_0$ with respect to our original frame. In this
coordinate system $(\varrho, \zeta, \varphi', t)$,
the only new coordinate reads
$ \varphi' = \varphi-\omega_0 t$.
Note that line element, Ernst equations and LP preserve their form in this rotating frame. The
transformed pseudopotential ${\bf\Omega}'$ can be obtained from
${\bf\Omega}$ via 
\begin{equation}\label{ansatz}
 {\bf\Omega}' = \left[\left(\begin{array}{ccc}
          c_- & 0    & 0\\
          0   & c_+  & 0\\
          0   & 0    & \sqrt{c_+ c_-}
          \end{array}\right)     
        +\alpha
          \left(\begin{array}{ccc}
          -1      & -\lambda & 0\\
          \lambda & 1        & 0\\
          0       & 0        & 0
          \end{array}\right)\right]{\bf\Omega}
\end{equation}
with 
\[
c_\pm := 1+\omega_0(a \pm \varrho\ee^{-2U}), \quad\alpha:=\ii(K+\ii z)\omega_0\ee^{-2U}.
\]
(Cf. \cite{Neugebauer4,Neugebauer2} for the corresponding transformation
valid in pure gravity, i.e., without electromagnetic field.) 

\vspace{-1ex}
\section{Solution of the linear problem\label{solution}}

We are able to integrate the LP along the dashed line in Fig.~\ref{fig1} (right
panel) since the metric inside the entire hyperbolic region
(excluding $\mathcal H^-$ for the time being) 
is regular. This regularity is a consequence of the requirement that the 
metric potentials be analytic functions of $R$ and $\cos\theta$
in an exterior vicinity of $\mathcal H^+$, including $\mathcal H^+$, see theorem 6.3
in \cite{Chrusciel1}. Note that with the arguments presented there, this theorem can be carried over
to the Einstein-Maxwell case considered here \cite{Chrusciel2}.

On the entire integration path, the Weyl coordinate
$\varrho$ vanishes, see \eqref{BoyerLcoord}, leading to
$\lambda=\pm 1$. For $\varrho=0$ and $\lambda=1$
the LP reduces to an ODE with the general solution
\begin{equation}\label{sol}
 {\bf\Omega} = \left(\begin{array}{ccc}
               \bar\E+2|\Phi|^2 & 1  & \Phi\\
               \E               & -1 & -\Phi\\
               -2\ii\ee^U\bar\Phi & 0 & -\ii\ee^U
               \end{array}\right){\bf C}(K).
\end{equation}
Here, ${\bf C}$ is a $3\times 3$ matrix that depends on $K$ only. 
Respecting the gauge \eqref{gauge}, the third column of ${\bf C}$ vanishes.

On all four parts of our integration path, ${\bf\Omega}$ has the form
$\eqref{sol}$, but with different ``integration constants'' ${\bf C}$.
We denote these with ${\bf C}$ on $\mathcal A^+$, $\tilde{\bf C}$ on
$\mathcal A^-$, ${\bf D}$ on $\mathcal H^+$, and $\tilde{\bf D}$ on
$\mathcal H^-$. Moreover, we can normalize ${\bf\Omega}$ such that 
\begin{equation}
 {\bf C}(K) = \left(\begin{array}{ccc}
               C_1(K) & 0 & 0\\
               C_2(K) & \psi(K) & 0\\
               C_3(K) & 0 & 0
               \end{array}\right),\ \psi:=(K^2-4\rh^2)^3,
\end{equation}
holds for the integration constant on $\mathcal A^+$.

From \eqref{sol}, we also calculate the pseudopotentials in the 
two different rotating frames of reference with $\omega_0=\omega^\pm$,
cf.~\eqref{ansatz}, where
$\omega^\pm=\omega|_{R=\pm\rh}=\textrm{constant}$
denotes the angular velocities of the horizons $\Hpm$.
In particular, $\omega^\pm$ and the metric potential $a$ are related
by the horizon boundary conditions 
$a=-1/\omega^\pm$ valid on $\mathcal H^\pm$.
Note that $a=0$ on $\mathcal A^\pm$.
Now, the pseudopotentials ${\bf\Omega}$ and ${\bf\Omega}'$ (in both rotating frames of reference) 
are continuous at the north and south poles of the horizons $\mathcal
H^\pm$. This leads us to an algebraic system of equations
for the elements of $\tilde{\bf C}$, ${\bf D}$ and $\tilde{\bf D}$
in terms of ${\bf C}$. From the solution of this system we are able to derive
the solution of the LP on the entire integration path, expressed in terms of the
three functions $C_1(K)$, $C_2(K)$, and $C_3(K)$.

\section{Potentials on the Cauchy horizon\label{sec:EP}}

From the pseudopotential ${\bf\Omega}$, we now calculate the
potentials $\E$ and $\Phi$ on $\mathcal H^-$. In a first step, we
express $C_1$, $C_2$, and $C_3$ in terms of the event horizon
potentials. 

At the branch points $K_1=\ii\bar z$ and $K_2=-\ii z$, ${\bf\Omega}$ is unique;
i.e., the values in both $K$-sheets coincide. 
In particular, for $\varrho=0$ (where $K_1=K_2=\zeta$) we have 
\begin{equation}\label{sheet}
 \psi_i^>=\psi_i^<, \quad i=1,2,3,\quad
 \textrm{for}\quad K=\zeta.
\end{equation}
Considering these conditions at $\mathcal H^+$, it follows that
\begin{subequations}\label{C}
\begin{eqnarray}\label{C1}
 C_1(\zeta) & = & n
       \left[\bar\E+2\Phi\pN\bar\Phi-2\ii\omega^+(\zeta-2\rh)+\E\pN\right],
       \qquad\\
 C_2(\zeta) & = & \psi+n 
   \left[(\E\pN+2|\Phi\pN|^2)(\bar\E+2\Phi\pN\bar\Phi+\E\pN)\right.\nonumber\\
 &&  \quad\qquad+\left.2\ii\omega^+(\zeta-2\rh)\bar\E\,\right],\\
 C_3(\zeta) & = & -2n
      \left[\bar\Phi\pN(\bar\E+2\Phi\pN\bar\Phi+\E\pN)\right.\nonumber\\
 &&    \quad\qquad-\left.2\ii\omega^+(\zeta-2\rh)\bar\Phi\,\right],
\end{eqnarray}
\end{subequations}
where $\E$ and $\Phi$ denote the potentials on $\mathcal H^+$ and
with
\begin{equation}
 n:=\frac{\E+2\bar\Phi\pN\Phi+2\ii\omega^+(\zeta-2\rh)-\E\pN-2|\Phi\pN|^2}
    {4(\omega^+)^2(\zeta-2\rh)^2\,\f}\psi.
\end{equation}
Here, we have introduced the notations $(\cdot)\pmN$ and $(\cdot)\pmS$ for
values at the north pole ($\theta=0$) and the south pole ($\theta=\pi$)
of the horizons $\mathcal H^\pm$ ($R=\pm\rh$) respectively.

Now, we evaluate \eqref{sheet} on $\mathcal H^-$ and
solve the resulting conditions for $\E$ and $\Phi$. In terms of
the Boyer-Lindquist-type coordinate $\theta$,
we obtain the remarkable explicit relations
\begin{subequations}\label{CHpots}
\begin{eqnarray*}
 \E^-(\theta) & = &
 \frac{a_1(\theta)\,\E^+(\pi-\theta)+a_2(\theta)\,\Phi^+(\pi-\theta)
 +a_3(\theta)} 
  {c_1(\theta)\,\E^+(\pi-\theta)+c_2(\theta)\,\Phi^+(\pi-\theta)+c_3(\theta)},
  \\ 
 \Phi^-(\theta) & = &
 \frac{b_1(\theta)\,\E^+(\pi-\theta)+b_2(\theta)\,\Phi^+(\pi-\theta)
 +b_3(\theta)} 
  {c_1(\theta)\,\E^+(\pi-\theta)+c_2(\theta)\,\Phi^+(\pi-\theta)+c_3(\theta)}, 
\end{eqnarray*}
\end{subequations}
between the potentials on $\mathcal H^-$ (superscript ``$-$'') and on
$\mathcal H^+$ (superscript ``$+$'').
The functions $a_i$, $b_i$, $c_i$, $i=1,2,3$, are specific polynomials in
$\cos\theta$ which will be given explicitly in \cite{Hennig4}.

A thorough discussion of \eqref{CHpots} reveals that the regularity of the potentials 
on $\mathcal H^+$ implies that of the potentials on $\mathcal H^-$, 
provided that $J$ and $Q$ do not both vanish.
In the limit of vanishing $J$ and $Q$,
the potentials $\E^-$ and $\Phi^-$ diverge.

\section{A universal equality\label{sec:eq}}

In this section, we derive that \eqref{ApAm} is true for arbitrary axisymmetric and
stationary black holes with surrounding matter in Einstein-Maxwell theory.
As in pure Einsteinian gravity (cf. \cite{Ansorg2}) 
it is possible to express $J$, $Q$, and $A^\pm$
in terms of potential values at north and south poles of the horizons $\Hpm$:
\begin{eqnarray*}
 J & = & \frac{1}{8(\omega^+)^2}\left[b\pS-b\pN-8\omega^+\rh
         +2A_t^+(\beta\pS-\beta\pN)\right],\\
 Q & = & \frac{1}{2\omega^+}(\beta\pN-\beta\pS),\quad
 A^\pm  =  \pm\frac{32\pi\rh}{\f_{,R}\big|\pmN},
\end{eqnarray*}
with $\omega^+ = \frac{1}{4}\left[b_{,R}+2(A_t\beta_{,R}-\beta
A_{t,R})\right]\pN.$

In order to calculate $\f_{,R}\big|\pmN$, we use the solution of the LP
on $\mathcal A^+$. Evaluation of the conditions \eqref{sheet}
leads us to the simple relation
\begin{equation}
\ee^{2U} = \frac{\psi(\zW)}{C_1(\zeta)}\quad\textrm{on}\quad \mathcal A^+.
\end{equation}
The explicit expression \eqref{C1} for $C_1(\zeta)$ and the potential horizon 
boundary conditions at $\mathcal H^+$, together with a careful study of the
Einstein-Maxwell equations at the north and south pole of $\mathcal H^+$, yields
\begin{subequations}\label{area}
\begin{eqnarray}
 A^+ & = & -\frac{2\pi}{(\omega^+)^2}\,\f_{,\theta\theta}\big|\pN\\
 A^- & = &  -\frac{\pi}{2(\omega^+)^2\,\f_{,\theta\theta}\big|\pN}
            \Big[(\beta\pN-\beta\pS)^4\nonumber\\
     &   & +\left.\left(b\pS-b\pN-8\omega^+\rh +2A_t^+
            (\beta\pS-\beta\pN)\right)^2\right].\qquad
\end{eqnarray}
\end{subequations}

Hence with the above expressions for $J,Q$ and $A^\pm$ we conclude that equation \eqref{ApAm} is satisfied. 

Finally, combining our results with a closely related inequality obtained in \cite{Hennig3},
we arrive at the following.
\begin{Thm}
	Every regular axisymmetric and stationary Einstein-Maxwell
	black hole with surrounding matter has a regular inner Cauchy
        horizon if and  
	only if the angular momentum $J$  and charge $Q$
        do not both vanish. Then the universal relation 
 	$$(8\pi J)^2+(4\pi Q^2)^2 = A^+ A^-$$
 	is satisfied where $A^+$ and $A^-$ denote the areas of event and
        inner Cauchy horizon 
 	respectively.
 	If, in addition, the black hole is sub-extremal (i.e., if there exist
 	trapped surfaces in every sufficiently small interior vicinity of the
	event horizon), then the following inequalities hold:
        $$A^-<\sqrt{(8\pi J)^2+(4\pi Q^2)^2}<A^+.$$
\end{Thm}

\begin{acknowledgments}
We would like to thank Gernot Neugebauer and
Piotr T. Chru\'sciel for many valuable discussions.
This work was supported by the Deutsche
Forschungsgemeinschaft (DFG) through the
Collaborative Research Centre SFB/TR7
``Gravitational wave astronomy''.
\end{acknowledgments}


\begin{thebibliography}{10}

\bibitem{Ansorg2}
M. Ansorg and J. Hennig,
Class. Quantum Grav. {\bf 25}, 222001 (2008).
\bibitem{Bardeen}
J. M. Bardeen, Rapidly rotating stars, disks, and black holes,
in {\it Black holes} (Les Houches), edited by C. deWitt and B. deWitt
(Gordon and Breach, London, 1973), pp. 241-289.
\bibitem{Carter}
B. Carter,
Black hole equilibrium states
in {\it Black Holes} (Les Houches),
edited by C. deWitt and B. deWitt
(Gordon and Breach, London, 1973), pp. 57-214.
\bibitem{Chrusciel1}
P. T. Chru\'sciel, Ann. Physics {\bf 202}, 100 (1990).
\bibitem{Chrusciel2}
P. T. Chru\'sciel, private communication. 
\bibitem{Ernst}
F. J. Ernst, Phys. Rev. {\bf 168}, 1415 (1968).
\bibitem{Hennig3}
J. Hennig, C. Cederbaum, and M. Ansorg,
submitted, arXiv:0812.2811.
\bibitem{Hennig4}
J. Hennig and M. Ansorg, arXiv:0904.2071.
\bibitem{Neugebauer}
G. Neugebauer and D. Kramer, J. Phys. A: Math. Gen. {\bf 16},
1927 (1983).
\bibitem{Neugebauer3}
G. Neugebauer and R. Meinel, Phys. Rev. Lett.
{\bf 73}, 2166 (1994); {\bf 75}, 3046 (1995).
\bibitem{Neugebauer4}
G. Neugebauer, Ann. Phys. (Leipzig) {\bf 9}, 342 (2000).
\bibitem{Neugebauer2}
G. Neugebauer and R. Meinel, J. Math. Phys. {\bf 44}, 3407 (2003).
\bibitem{ExactSolutions}
H. Stephani, D. Kramer, M. MacCallum, C. Hoenselaers, and E. Herlt,
{\it Exact Solutions of Einstein's Field Equations}
(University Press, Cambridge, 2003).

\end{thebibliography}
\end{document}